\documentclass[pre,preprint,aps,]{revtex4}
\usepackage{graphicx}
\usepackage{comment}
\usepackage{amsmath,amssymb,amsthm} 
\usepackage{bm}
\usepackage{verbatim}
\usepackage{float}

\begin{document}

\title{An algorithm for coalescence of classical objects and formation of planetary systems}




\author{ S\o ren Toxvaerd}
%
%
\affiliation{ Department
 of Science and Environment, Roskilde University, Postbox 260, DK-4000 Roskilde, Denmark}
 \date{\today}

 \vspace*{0.7cm}

 \begin{abstract}

Isaac Newton formulated the central difference algorithm (Eur. Phys. J.  Plus (2020) 135:267) when he   derived his second law.
 The algorithm is under  various names (''Verlet, leap-frog,..") the most used algorithm in simulations of complex system in Physics and Chemistry, and it
 is also applied in Astrophysics.
 His discrete dynamics has the same qualities as his exact analytic dynamics for continuous space and time with time reversibility, symplecticity and conservation of
 momentum, angular momentum and energy. Here the algorithm is extended to include the fusion of objects at collisions.
 The extended algorithm is used to
 obtain the self-assembly of celestial objects at the emergence of planetary systems.  The emergence of twelve planetary systems is obtained.
 The systems are stable over very long times, even when two ''planets" collide  or if a planet is engulfed by its sun.
 \end{abstract}

\maketitle

\section{Introduction}

 Simulations of collections of classical objects in the Universe have been performed for many decades
 \cite{Klypin1983,Centrella1983,Wisdom1991,Duncan1998,Chambers1998,Robutel2001,Reintamayo2015,Rein2019}. 
 The coupled second order differential equations have either been solved  numerically by various 
 higher- order symplectic  algorithms \cite{Wisdom1991,Forest1990,Yoshida1990,Hernandez2015,petit2019} or  
  by the  Particle-Particle/Particle-Mess (PPPM)  method  \cite{Hockney1974}.  
 In the PPPM method  each mass unit, e.g. a planet is treated as  moving in the collective field of all others, and the  Poisson equation
 for the PPPM grid is solved numerically.
 Later, simulations with  large scaled computer packages \cite{Diemand2008,Alimi2012} with  many billions of  mass units show strong evidence
 for dark matter in the Universe. A comprehensive review of the  simulations, the mass distribution in the Universe and
 the evidence for dark matter is given in \cite{Martino2020}.

The present  algorithm and the simulations deviate from the  previous algorithms and the  PPPM model and the complex simulations in several ways.
The basic algorithm is the central difference algorithm.
It appears in the literature under different names, most known as the ''Verlet-" or ''leap-frog" algorithm,
but it is actually first formulated by Isaac Newton  in  PHILOSOPHI\AE \ NATURALIS PRINCIPIA MATHEMATICA in 1687  \cite{Newton1687,Newton1}.
In celestial mechanics it has been rediscovered  as the second-order leap-frog discrete mapping
\cite{Wisdom1991,Forest1990,Yoshida1990} and extended to higher order \cite {Hernandez2015,petit2019}. 
Newtons discrete expression for the
relation between the positions of the objects, their forces and the discrete time propagation is time reversible, symplectic and has the same dynamic  invariances for
  a conservative system as his analytic formulation \cite{Tox1,Toxa}. The algorithm allows for obtaining the
 discrete dynamics of classical objects without  any approximations, and here his algorithm is extended to  cover
   the fusions of objects and self-assembly at the emergence  of planetary systems. 

\section{The discrete algorithm for fusion of classical objects}
According to Newton's classical discrete dynamics  a new  position $\textbf{r}_k(t+\delta t)$ at time $t+\delta t$ of an object
$k$ with the mass $m_k$  is determined by
the force $\textbf{f}_k(t)$ acting on the object   at the discrete position $\textbf{r}_k(t)$  at time $t$ and the position 
$\textbf{r}_k(t-\delta t)$ at $t - \delta t$  as
\begin{equation}
	 m_k\frac{\textbf{r}_k(t+\delta t)-\textbf{r}_k(t)}{\delta t}
			=m_k\frac{\textbf{r}_k(t)-\textbf{r}_k(t-\delta t)}{\delta t} +\delta t \textbf{f}_k(t).	
 \end{equation}
where the momenta $ \textbf{p}_k(t+\delta t/2) =  m_k (\textbf{r}_k(t+\delta t)-\textbf{r}_k(t))/\delta t$ and
 $  \textbf{p}_k(t-\delta t/2)=  m_k(\textbf{r}_k(t)-\textbf{r}_k(t-\delta t))/\delta t$ are constant in
 the time intervals in between the discrete positions.
 Newton  $postulated$ Eq. (1) and obtained his  second  law from Eq. (1)  as the limit $ lim_{\delta t \rightarrow 0}$ \cite{Newton1}.

 Newton is together with Leibniz the fathers of analytic mathematics and Newton's discrete algorithm- or equivalent expressions, is usually presented as a third-order predictor algorithm, which can be derived 
 by a Taylor-McLaurin expansion from the objects analytic trajectories. Brok Taylor (1685-1731) lived at the same time as Newton (1643-1727), and Newton had full knowledge of Taylor expansions,
 but Newton  never presented his 
 expression for his second law, even in his later two editions of $Principia$, as the first and leading term in an analytic expansion. And with good reason because,  unlike algorithms obtained by higher-order 
 expansions, his discrete algorithm has all the qualities of the analytic analog.

 Isaac Newton obtained his second law as the limit expression   $ lim_{\delta t \rightarrow 0}$ of the central  difference 
   in momentum for a planet at a  discrete change  $\delta t$ in time. But he noticed in $Principia$ at the derivation of the law,
   that the areas of three triangles in his geometrical construction
   of the discrete trajectory of a planet are equal, \textit{an irrelevant observation for the derivation},
   but he did not mentioned the consequence of the equal areas.
   It is Kepler's second law, and the young  Newton must immediately, when he postulated the law have realised,
   that his discrete relation Eq. (1) at least explains Kepler's second law. But he 
   did not mentioned it at the derivation of the second law, even much later when he wrote
   $Principia$, nor in his  second- or third editions of $Principia$  \cite{Newton1687}.
    The fulfillment of  Kepler's second law is a consequence of the conserved angular momentum in his discrete dynamics \cite{Tox3}. An explanation for that Newton on one hand
     noticed the equality of the areas of the triangles, and on the other hand did not noticed  that this explains Kepler's second law could be, that
      he believed that the exact classical dynamics first is achieved in the analytic limit with continuous time and space. But this is in fact not the
       case, his discrete dynamics has the same invariances as his analytic dynamics.
 Due to the time symmetry it s time reversible and the algorithm is also  symplectic \cite{Wisdom1991,Forest1990,Yoshida1990,Tox4}. The conservation of momentum and angular momentum
       is ensured by Newton's second and third law, because the sum of the forces between the objects
       in a conservative system is zero. The algorithm conserves also the energy, but  it is, however, not obvious because of the asynchronous 
       appearance of positions and momenta, and thereby the asynchronous determination of the
       potential- and the kinetic energy. But one can prove that the discrete algorithm conserves the energy \cite{Toxa} and  also show that there (most likely) exists a  ''shadow Hamiltonian" nearby
       the Hamiltonian for the analytic dynamics and where the discrete positions
       are located on the shadow Hamiltonian's analytic trajectories \cite{Tox1,Tox2,Tox3}. So the discrete dynamics has a constant energy given by the energy of the shadow Hamiltonian.

       Newton's discrete algorithm has been rediscovered several times, most known by L. Verlet \cite{Verlet}, and it appears with a variety of names:
       Verlet-, Leap-frog,...\cite{Tox3}.  Almost all Molecular Dynamics (MD) simulations of complex physical and chemical systems and many celestial mechanics simulations are performed with Newton's discrete algorithm.

It is convenient to reformulate Newton's algorithm  as the ''Leap frog" algorithm
\begin{equation}
\textbf{v}_k(t+\delta t/2)=  \textbf{v}_k(t-\delta t/2)+ \delta t/m_k  \textbf{f}_k(t),
\end{equation}
with the velocities $\textbf{v}(t+\delta t/2)$ and  $\textbf{v}(t-\delta t/2)$ and  the positions 
\begin{equation}
\textbf{r}_k(t+\delta t)= \textbf{r}_k(t)+ \delta t \textbf{v}_k(t+\delta t/2),	  
\end{equation}	  
so the new positions at $t+\delta t$ are obtained in two steps, first  by calculating the new
 (mean) velocities $ \textbf{v}_k(t+\delta t/2)$ in the time interval $[t,t+\delta t] $ from the old velocities $  \textbf{v}_k(t-\delta t/2)$
 in the previous time interval and the
forces  $\textbf{f}_k(\textbf{r}_k(t))$ at the positions $\textbf{r}_k(t)$, and then  the new
positions $\textbf{r}_k(t+\delta t)$ are obtained from the velocities  $ \textbf{v}_k(t+\delta t/2)$ .

 Newton's discrete algorithm is here used as a starting point  for a formulation of a discrete  algorithm
 for the irreversible fusion of spherical symmetrical objects 
 by classical dynamics  with inelastic collisions. The derivation of the algorithm is governed by a desire to preserve
 as much as possible of the invariances of Newton's dynamics.

\subsection{An algorithm for coalescence of classical objects and formation of planetary systems}
  The classical discrete dynamics between $N$ spherically symmetrical objects
     with masses $ m^N=m_1, m_2,..,m_N$ and positions $\textbf{r}^N(t)=\textbf{r}_1, \textbf{r}_2,..,\textbf{r}_N$  is obtained 
 by Eq. (2) and Eq.(3) with extensions.

According to Newton's shell theorem \cite{Newtonshell} the force, $\textbf{F}_i$, 
on  a spherically symmetrical object $i$ with mass $m_i$ is a sum over the forces, $ \textbf{f}(r_{ij})$, caused by the other 
spherically symmetrical objects $j$ with mass $m_j$, and it
is  solely  given by their center of mass distance $r_{ij}$ to $i$
 \begin{equation}
	 \textbf{F}_i(r_{ij})= \Sigma_{j \neq i}^N \textbf{f}(r_{ij})= -\frac{G m_i m_j}{r_{ij}^2}\hat{\textbf{r}}_{ij}.
 \end{equation}

Let all the spherically symmetrical objects
 have the same (reduced)  number density $\rho= (\pi/6)^{-1} $ by which
 the diameter $\sigma_i$ of the spherical object $i$ is 

 \begin{equation}
	 		 \sigma_i= m_i^{1/3}
 \end{equation}
  and   the collision diameter 
\begin{equation}
	\sigma_{ij}=	\frac{\sigma_{i}+\sigma_{j}}{2}.
\end{equation}	
 If  the distance $r_{ij}(t)$ at time $t$ between two objects is less than $\sigma_{ij}$ 
the two objects merge to one spherical symmetrical object with mass

\begin{equation}
m_{\alpha}= m_i + m_j,
\end{equation}	 
and diameter
\begin{equation}
 \sigma_{\alpha}= (m_{\alpha})^{1/3},
\end{equation}
and with the new object $\alpha$  at the position
\begin{equation}
	\textbf{r}_{\alpha}= \frac{m_i}{m_{\alpha}}\textbf{r}_i+\frac{m_j}{m_{\alpha}}\textbf{r}_j,
\end{equation}	
at the center of mass of the the two objects before the fusion.
(The   object $\alpha$ at the center of mass of the two merged objects $i$ and $j$ might occasionally be near another object $k$
by which more objects merge, but after the same laws.)

Let the center of mass of the system of  the $N$ objects be at the origin, i.e.
\begin{equation}
			\Sigma_k m_k \textbf{r}_k(t)=\textbf{0}.
\end{equation}

The momenta  of the objects in the discrete dynamics just before the fusion are $\textbf{p}^N(t-\delta t/2)$ and the
total momentum of the system is conserved  at the fusion if
\begin{equation}
\textbf{v}_{\alpha}(t-\delta t/2)= \frac{m_i}{m_{\alpha}}\textbf{v}_i(t-\delta t/2)+ \frac{m_j}{m_{\alpha}}\textbf{v}_j(t-\delta t/2),
\end{equation}
which determines the  velocity $\textbf{v}_{\alpha}(t-\delta t/2)$ of the merged object.

The invariances in the classical
Newtonian dynamics are for a conservative system with Newton's third law, i.e with
\begin{equation}
	\textbf{f}_{kl}(t)=-\textbf{f}_{lk}(t)
\end{equation}	
for the forces between two objects $k$ and $l$, and with no external forces.
An  object $k$'s forces with $i$ and $j$ before the fusion 
 are   $\textbf{f}_{ik}(t)$ and $\textbf{f}_{jk}(t)$,
and these forces
must be replaced by calculating the force $\textbf{f}_{\alpha k}(\textbf{r}_{\alpha k}(t))$.
The total force after the fusion is zero 
due to Newtons third law  for a conservative system with  the forces $\textbf{f}_{\alpha k}=-\textbf{f}_{k \alpha}$ between pairs of objects,
 and the total momentum
\begin{eqnarray}
	\Sigma_k \textbf{p}_k(t_n+\delta t/2)= \Sigma_k \textbf{p}_k(t_n-\delta t/2)+ \delta t\Sigma_k \textbf{f}_k(t_{n}) \nonumber \\
	= \Sigma_k \textbf{p}_k(t_n-\delta t/2),
\end{eqnarray}	
and the  position of the center of mass are conserved for the discrete dynamics with fusion.

 The determination of the position, $\textbf{r}_\alpha(t)$, and 
 the velocity,  $\textbf{v}_\alpha(t-\delta t/2)$,  of the new object from the requirement of  
 conserved center of mass and  conserved momentum determines  the discrete dynamics of the $N-1$ objects.

 The angular momentum is   affected by the fusion.
 The angular momentum of the system of  spherically symmetrical objects consist of two terms
 \begin{equation} 
	 \textbf{L}(t)= \textbf{L}_{G}(t)+ \textbf{L}_{I}(t)
 \end{equation}
 where $ \textbf{L}_{G}(t)$ is the angular momentum of the  objects due to the dynamics obtained from the gravitational forces between their
 center of masses, and $\textbf{L}_{I}(t)$ is the angular momentum  due to the  spin of the objects. 
 Without fusion  $\textbf{L}_{G}(t)$    is  conserved for Newtons discrete
dynamics \cite{Tox3}.    $\textbf{L}_{I}(t)$ is, however, also  conserved  according  to the shell theorem \cite{Newtonshell} , where Newton
proves that  no net gravitational force is exerted by
a shell on any object inside, regardless of the object's location within the uniform shell, by which the
spin of the object is not affected by any force and is therefore constant. 
But at a fusion $ \textbf{L}_{G}$ changes  by
\begin{equation}
	\delta \textbf{L}_{G}(t)= \textbf{r}_{\alpha}(t) \times m_{\alpha}\textbf{v}_{\alpha}(t-\delta t/2)-
 \textbf{r}_i(t) \times m_i\textbf{v}_i(t-\delta t/2)- \textbf{r}_j(t) \times m_j\textbf{v}_j(t-\delta t/2).
\end{equation}	 
and $ \textbf{L}_{I}$ changes  by
\begin{eqnarray}
\delta \textbf{L}_{I}(t)=
(\textbf{r}_i(t)- \textbf{r}_{\alpha}(t))\times m_i\textbf{v}_i(t-\delta t/2)+
(\textbf{r}_j(t)- \textbf{r}_{\alpha}(t)) \times m_j\textbf{v}_j(t-\delta t/2) \nonumber \\ \nonumber
= \textbf{r}_i(t) \times m_i\textbf{v}_i(t-\delta t/2)+ \textbf{r}_j(t) \times m_j\textbf{v}_j(t-\delta t/2)
 - \textbf{r}_{\alpha}(t) \times m_{\alpha}\textbf{v}_{\alpha}(t-\delta t/2)  \\
=	-\delta \textbf{L}_{G}(t).	
\end{eqnarray}	 

So  without fusion the angular momenta $ \textbf{L}_{I}(t)$ and $ \textbf{L}_{G}(t)$ with Newton's discrete dynamics  are
conserved separately, and at a fusion the  total angular momentum is still  conserved but with an exchange of angular momentum with
$\delta \textbf{L}_{I}(t)= -\delta \textbf{L}_{G}(t)$.

The exact classical discrete dynamics with fusion of colliding objects can be used to explore the self-assembly at the emergence of planetary systems  and to investigate
the stability and chaotic behaviour  of solar systems \cite{Hernandez2020}.

\begin{figure}
	\begin{center}
	\resizebox{0.5\textwidth}{!}{		
	\includegraphics[angle=-90]{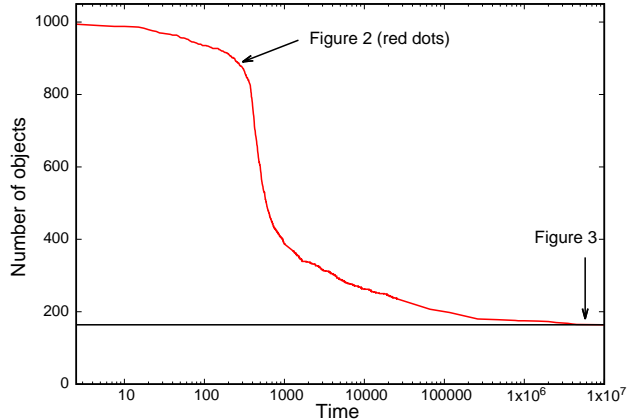}}
	\caption{ The number $N(t)$ of objects (sun, planets and free objects) as a function of time $t$ with fusion for one (No. 1) of the twelve systems.
  This  system contained 852 objects at  $t=250$.  The
positions of the $N$=852 objects are shown in Figure 2  (red dots) together with the
	start positions with  $N=1000$ (small blue dots). At $t= 4.5\times 10^6$ the planetary system contained one Sun
		and 165 planets and free objects, and  the system aged with only one fusion for 
		the succeeding $\Delta t=5.5 \times 10^6$ time.   The planet orbits at  $t=4.5 \times 10^6$ 
		for four inner planets are shown in Figure 3.}
	\end{center}
\end{figure}

\begin{figure}
	\begin{center}
	\resizebox{0.65\textwidth}{!}{
		\includegraphics[angle=-90]{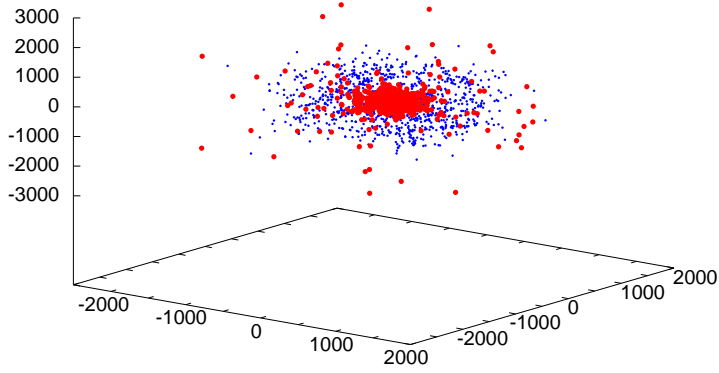}}

		\caption{ The positions of the $N$ objects at the start
		of fusion with small blue dots, and  with red dots 	at $t=250$ where   the fusion accelerated (Figure 1) and ended with   
	 	 one sun, 23 planets and 142 free objects.}
	\end{center}
\end{figure}

\begin{figure}
	\begin{center}
	\resizebox{0.65\textwidth}{!}{
		\includegraphics[angle=-90]{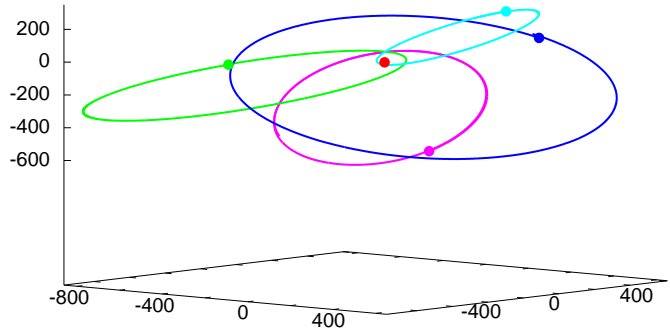}}
	\caption{ The sun (red, enlarged) with  four  planets close to the sun.
	The orbits are obtained  at $t=4.50 \times 10^6$   after 
	creation of the solar system. Light blue: Orbit time
		$T_{\textrm{orbit}}=630,$  eccentricity $ \epsilon=0.941$; green: $T_{\textrm{orbit}}=1308, \epsilon=0.867$;
	blue	$T_{\textrm{orbit}}=1740, \epsilon=0.377$;
magenta $T_{\textrm{orbit}}=1529, \epsilon=0.815$. The light blue planet have circulated $\approx$ seven thousand times around
the sun.}
	\end{center}
\end{figure}

 \begin{figure}
 \begin{center}
 \resizebox{0.5\textwidth}{!}{\includegraphics[angle=-90]{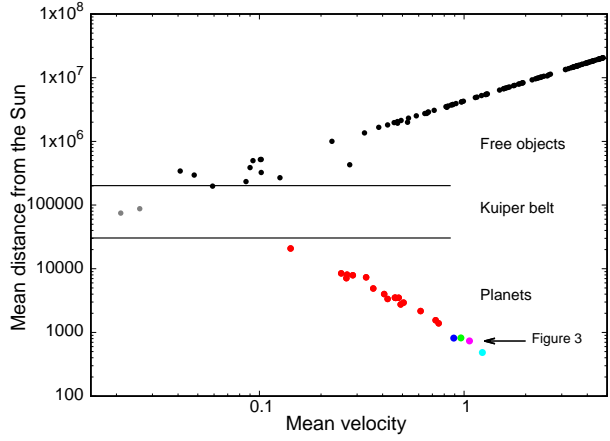}}
 \caption{ Mean $log-$distances ($log <r_{i,\textrm{sun}}(t)>$)
 to the  sun of the objects $i$  as a function of their relative  ($log$) mean velocities 
 $log(<v_{i,\textrm{sun}}(t))>$) for planetary system No. 1. The means are for a time interval
 $\Delta t \in  [4.0 \times 10^6, 4.5 \times 10^6]$. The locations of the four inner planets in Figure 3 are marked with their color from Figure 3.
 The planets (colored spheres) are located on the lower branch of the distribution and the upper
 branch shows(black spheres) the mean locations of the free objects. The ''Kuiper belt" with there objects (grey spheres) is estimated to
 be for mean locations $\approx 
 <r_{i,\textrm{sun}}(t)> \in [30000,200 000]$. Orbits of  planets  in the Kuiper belt are shown in the next figure.}
 \end{center}		
 \end{figure}

\begin{figure}
\begin{center}
\resizebox{0.65\textwidth}{!}{\includegraphics[angle=-90]{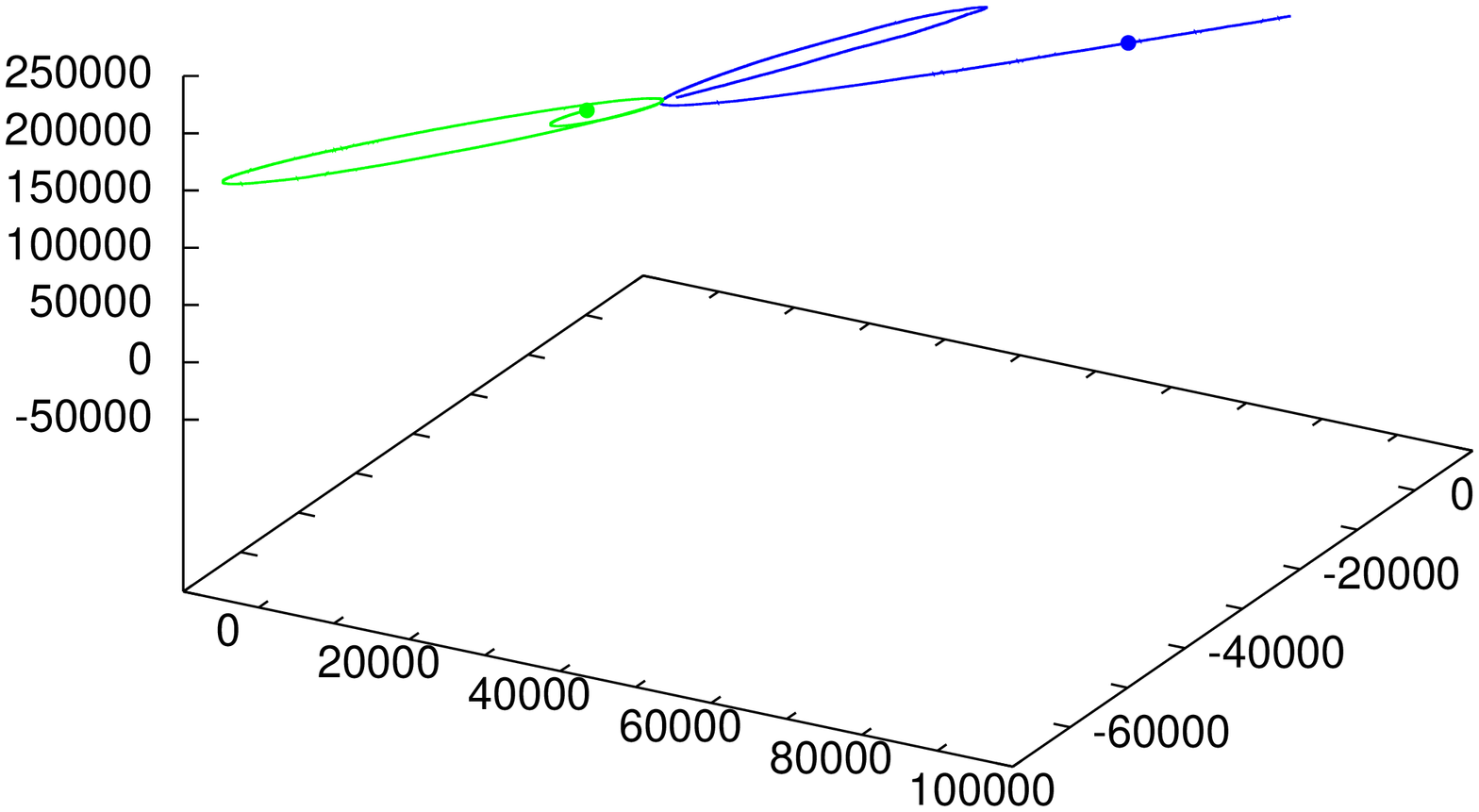}}
\caption{ Two planets in the ''Kuiper belt". The planet shown with green changed its course, but remained in the Kuiper belt,
	whereas the planet shown with blue escaped
	the planetary system.}
\end{center}		
\end{figure}

\section{Simulation of  formation   of  planetary systems }

The algorithm Eqs.(2), (3) and IIA is used to simulate the emergence of planetary systems. The  set-up of the actual MD systems and the general conditions
for MD for gravitational systems  is given in  the  Appendix.
Our planetary system is presumably created   from flattened, rotationally supported disc structures of cosmic dust grains, and the composition of the building blocks -
 planetesimals - is grossly different from that of the sun \cite{Blum2008}. 
  Here the 
results for twelve MD simulations  of the emergence of  planetary systems  are presented.
The planetary systems are obtained from different diluted ''gas" states of $N=1000$ objects with equal masses and at different low
temperatures (mean velocity of the objects), and the systems can be  considered as 
embryos of planetary systems by self-assembly of simple  small grains.

The creation of a  system with one heavy central object and with some of the other objects in orbits around the central ''sun" is
established within a relative short period of time as illustrated in Figure 1. The
start configurations are  diluted spherical (gas) distributions of objects (see Appendix).
The objects are accelerated toward the center of mass by the gravitational forces,
and the  fusion of objects results in a creation of a system with one heavy object (the sun) and
with other of  the objects in elliptical orbits around the sun. The solar systems are created rather quickly.
The system No. 1 (Figure 1-5,  and Table I) is established  already after a fusion time  $t \approx 1000$ (see Figure 1, for MD details and unit of time see Appendix)
with 386 objects consisting of one sun  with the mass $m_{\textrm{sun}}=557$ and many planets  and free objects.
Nine of the planets have a mass $m=3$, but most  of the other planets and free objects (338)
 are not fused with others and have a mass $m=$ 1. The solar system is in a rather stable state but  ages slowly (Figure 1). First after  $t=4.5\times 10^6$ are  all
 twelve planetary system stable and with very rare mergers (Table I).
The four planets in planetary system No. 1  closest to the sun is shown in Figure 4. The  angular momentum, $\textbf{L}_G$,
of the planetary system is constant if there
is no fusion, but also the angular momenta of the individual planets are also almost constant.
The four planets shown in Figure 4 have    almost constant angular momenta with closed elliptical orbits.

The solar system shown in the Figures 1-4 is  system No. 1 (see Table I). It consists of many planets including very
tiny bounded planets in a ''Kuiper belt'' and with orbits which have an orbit time  of more than
$t_{\textrm{orbit}}=1 \times 10^5$. The existence of such
Kuiper belt makes it difficult to determine precisely how many planets a planetary system consists of since the planets change their
orbits over time. The distribution of planets and free objects is determined from the distances and velocities
relative to the sun.  A  planet will have a relative short mean distance to the sun, averaged
over a long time interval, whereas a free object has a long  mean distance and a constant velocity. Figure
4 shows the relative mean distances  of the objects to the sun as a function of their relative  mean velocities, where the means are obtained
for $\Delta t \in [4.0\times 10^6,4.5\times 10^6]$.
The distribution has two branches, a lover
branch for the planets and an upper branch for the free objects.

The Kuiper belt is located  in between the two branches in Figure 4.
The objects in this zone  are almost free from the gravitational attractions of the sun and the other planets, and sometimes an object
in this zone escapes from the
planetary system. Figure 5 shows two planets in the Kuiper belt, where one (green) remained in the planetary system,
whereas one (blue) escaped. The planet with green remained in
the planetary system and with  elliptical-like orbits.
 The   planet with green was in an elliptical orbit with an eccentricity 
\begin{equation}
\epsilon =\frac{r_{\textrm{max}} -r_{\textrm{min}}}{r_{\textrm{max}} +r_{\textrm{min}}}=\frac{156090-115}{156090+115}=0.9985,	
 \end{equation}	
and  with the longest distance $r_{\textrm{max}}=156090$ at aphelion and the shortest distance at perihelion 
 $r_{\textrm{min}}=115$. The orbit time is $5.06 \times 10^5$. After passing the aphelion at $r_{i,sun}$=156090 the planet ended in
 a new elliptical orbit closer to the sun with a new aphelion distance  $r_{i,sun}$=38635. The other object shown with blue in Figure 5 excaped the planetary system. 

Our Solar system has a Kuiper belt located $\approx$ 30-100 AU (astronomical unit = mean distance between the Earth and the Sun).
 This distances is translated to the present solar systems by setting 1 AU=500, i.e $\approx$  equal to the mean distance of one of the
 inner planets  in Figure 4 (light blue). 
 The  lower border  of the present Kuiper belt   should be $\approx 30 \times 500=15000$ with this unit  and the upper border 
 should be $\approx 50000$.  The 
 planet orbit  shown with green  in Figure 5 have a  maximum distance  156090.
 The present Kuiper belt in planetary system No. 1
is estimated to be in the interval 
 $\mid \textbf{r}(t)-\textbf{r}_{\textrm{sun}}(t) \mid \in [30000,200000]$.

The data for the twelve simulations of planets systems are collected in Table I.
 The data   are obtained in the  
 time interval $t \in [4.0\times 10^6, 4.50\times 10^6 ]$ after the start
 of the fusions. The mean distances $ <r_{\textrm{cl}}>$ is the mean distance to the sun
 for the planet in a system closest to the sun. The temperatures, $T$, of the objects are obtained from the
 mean kinetic energy $<E_{\small{\textrm{Kin}}}>/N=3/2T$, and the different start distributions and kinetic energies  of the objects result in  temperatures
 which varies with a factor of $\approx 4$. There is, however, no clear  connection between the number of planets and the mean kinetic energies
 of the planetary systems.
 
 The eccentricities and mean positions of planets in the twelve systems are shown in Figure 6.
 The distributions show that the inner planets in general have an eccentricity significant
 below $\epsilon=1$, which is the limit of stability for an  elliptical orbit, whereas the planets close to the border of the  the Kuiper belt ($\approx 30000$)
 all have eccentricities only little less than  the  limit of stability.

	 	 \textbf{Table 1}. Collected data for the  planetary systems
		  \begin{tabbing}
\hspace{1.6cm}\=\hspace{1.5cm}\=\hspace{1.6cm}\=\hspace{1.6cm}\=\hspace{1.6cm}\=\hspace{1.6cm}\=\hspace{2.0cm} \\
	\> Inner \>''Kuiper  \> Free  \\			  
  No.	\>Planets \> planets \> objects \> $m_{\textrm{Sun}}$\>$<r_{\textrm{cl}}>$\> Temp.  \\
--------------------------------------------------------------------------------------------\\
  1    \>  21 \> 2 \> 142\> 830 \> 484\> 4.24   \\
  2    \>  25 \> 10 \> 317 \> 628 \>119  \> 8.88   \\	  
  3    \>  24 \> 10 \> 229\> 720 \> 150\> 11.60  \\  
  4    \> 13   \>5 \> 224  \> 739 \> 455\> 12.53  \\
  5    \>  23 \> 7 \> 253\> 700 \> 191\> 10.76  \\
 6    \> 8  \>3 \> 116 \> 865\> 131\> 5.00  \\
  7    \>  7 \> 0\> 147 \> 839\>262 \>19.08   \\
 8    \> 13   \>4 \> 182 \> 765 \> 452   \> 3.45 \\
 9    \> 15  \>6 \> 159 \> 779 \>402  \> 4.62  \\
 10    \> 12   \>2 \> 141 \> 826 \>44 \> 3.95  \\
 11    \> 8   \>2 \> 107 \> 862  \>43 \> 4.76 \\
 12    \> 6    \>2 \> 90 \> 897 \>145 \> 4.65 \\

\end{tabbing}
--------------------------------------------------------------------------------------------\\

\begin{figure}
	\begin{center}
		\resizebox{0.5\textwidth}{!}{\includegraphics[angle=-90]{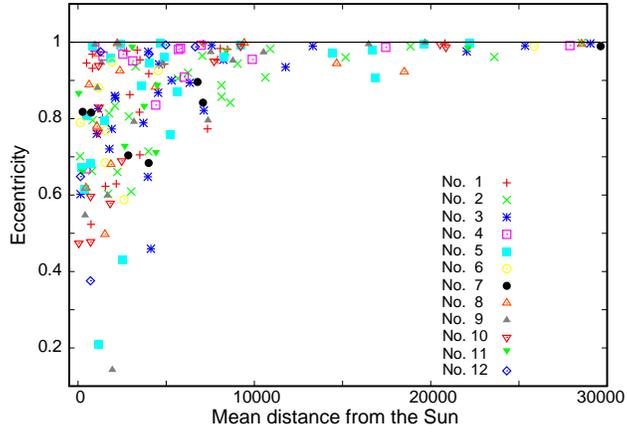}}
		\caption{   The eccentricity of the planets in the twelve planetary systems as
		a function of their mean distances from their suns.}
	\end{center}		
\end{figure} 

\section{Conclusion}

 Newton solved in $Principia$  as the first, Kepler's equation and determined the analytic expression for  the orbit of a planet -or a  comet,
 but  the  analytic dynamics of a solar system with many
 planets can only be obtained numerically, traditionally by the use of higher order symplectic algorithms.
  But the discrete dynamics with Newtons  central difference algorithm  is also time reversible and  symplectic and has the same invariances as his analytic classical dynamics.
 Here his discrete algorithm is extended to handle 
 fusion of objects at a collision and to create small planetary systems.
 
 The formation of a planetary system depends on the distribution and the kinetic energies of the collection of objects that start to fuse together.
 Our planetary system is presumably created   from flattened, rotationally supported disc structures of cosmic dust grains, and the composition of the building blocks -
 planetesimals - is grossly different from that of the sun \cite{Blum2008}.  
 Here  twelve  planetary systems are created  by fusions of  a small number $N=1000$ objects (planetesimals)
 with equal masses  and from a spherical starting distribution of the merging objects, in order to test the algorithm and to obtain an ''embryo" of a planetary system.
 The time evolutions in the  systems reveal that the objects spontaneously form ''mini" planetary systems with a heavy ''sun" and with many of the planetesimals in orbits
 around the sun.
 The planetary systems have some qualities which agree with our own Solar system, with stable elliptical orbits and with
 many bounded objects at great distances from the sun in a ''Kuiper belt". But the planetary systems deviate from the Solar system by,
 that the orbits are not in  a common  Ecliptic plane due to the spherically distributed starting positions  of the  merging objects. Furthermore there is no planetary systems with  moons. These deviations can, however, very well be
 a consequence of the small size of the systems with only $N=1000$ spherically distributed  objects at the start of the fusion, and to the monodisperticity of the
 systems with  equal masses of the  objects at the start. The  small systems   was selected in order to be able to follow
 the created solar systems over very long times without any approximations in order to test the exact algorithm and the aging and  stability of the planetary systems.
 
 The planetary systems are established over a short period of time with fusions.
 The systems are stable and age slowly by that  a planet occasionally collided with another planet or with the sun and merge. Some of the
 planets  were also accelerated out of the planetary system (Figure 5). The planetary systems show chaotic sensitivity, and the actual numbers of inner planets and their
 positions and eccentricities depend on the forces from all the other objects in the system, including the free objects far from the sun. Almost all of the
 twelve planetary systems
 contain many planets in  Kuiper belts far from the suns. 

 The extension of Newton's discrete dynamics  with the algorithm for fusion of colliding objects is the simplest possible. The fusion of two
 spherical symmetrical objects to one uniform and spherical symmetrical object is far from what actually happens when two macroscopic celestial bodies
 merge \cite{Canup2001}. But, although it is straight forward to extend the algorithm to  a more complex fusion at the  collision, it has not be the goal
 with the present investigation. The algorithm is suitable for analysis of the self-assembly of planetesimals and, due to the exact dynamics, the algorithm can be
 useful at  investigations of the impact of the chaotic behaviour  on the stability of planetary systems.
\\
$ $\\
$\textbf{Acknowledgement}$
This work was supported by the VILLUM Foundation Matter project, grant No. 16515.
\\
$\textbf{Data Availability Statement}$ Data will be available on request.

\section{Appendix Molecular Dynamics simulation of  planetary systems}

The simulations of the creation of planetary systems were performed for different numbers $N=100, 1000$ and $10000$ of objects and for
different start configurations of the objects. The spherically symmetrical objects are identical and interact with the gravitational force Eq. (4).
All units (mass, time, length, energy/force) are given in units given by $G, m_i$ and $\sigma_i$, and
the present simulations are started with $N=1000$ objects with equal masses
$m_i=1$ and diameter $\sigma_i=1$. In order to simulate the orbit of a planet near the sun it is necessary to
choose a small time increment $\delta t=0.0025$ \cite{Tox3}.
 The MD simulations are  with 
 double-precision variables, the center of mass and the momentum and orbits of planets are conserved after more than $10^9$ time steps with
 a small time increment $\delta t=0.0025$.

The phase space  diagram for the gravitational system \cite{Hernandez} differs from a traditional phase space  diagram. 
A collection of objects will, without fusion collapse  at low temperature (velocities of the objects) and relative
high concentration in the collection \cite{Lindblad}, but at high temperatures and low concentrations the 
collection of objects expands continuously in the space. The collapsed spherical symmetrical objects will, without fusion perform a crystal.
It is as mentioned  not possible to obtain a traditional phase diagram for a classical
system with gravitational forces
 because the  energy per object
\begin{eqnarray}
		u/kT=2 \pi \rho \int_0^{\infty} u(r) g(r) r^2 dr \nonumber \\
			\approx -2 \pi \rho G m_1 m_2 \int^\infty r_{12} dr_{12}
\end{eqnarray}	
(and free energy) diverges for an uniform  distribution. 
The objects crystallize at low temperatures if the objects perform elastic collisions without fusion.
  In contrast to this behavior the collections of free objects at low temperatures
  and concentrations and with  fusion merge into solar systems with planets in regular orbits. 
  The algorithm with fusions is used obtain  planetary systems.

Here we shall describe twelve simulations of the emergence of planetary systems for $N=1000$ and for  diluted gas configurations of the
objects at the start, which are spherically (blue dots in Figure 3). The twelve systems are generated for different start configurations and kinetic energies.  The momenta and angular momenta $\textbf{L}_G(0)$ at the start
$t=0$ are adjusted to zero.

Depending on the start configurations of positions and kinetic energy,  the systems either collapse at low kinetic energy and high
gas density into one heavy object (black hole), or expand  in the open space for high kinetic energy and low density. But in between these
ranges of velocities and concentrations some objects merge with creation of one heavy object (the sun), with planets and with unbounded objects.
(A free object is characterized by the fact that it, with a constant direction with respect to the  the sun 
and with a positive energy $E=E_{\textrm{Kin}}+E_{\textrm{Pot}}$, increases its distance
to the sun.)

The twelve systems are started at different  mean ''temperatures" $T$ in the interval $ T \in [0.1,0.5]$. The fusions started shortly
after (Figure 2) with an increase of the mean velocities. The total momenta and the centre of masses and angular momenta $\textbf{L}$ are conserved,
but the angular momenta  $\textbf{L}_G$
are not conserved at the  fusions, but varied within the range $\textbf{L}_G \approx  [\textbf{-10},\textbf{-10}]$.
The random round-off  errors in the double precision
arithmetic have no effect on the orbits of the planets. The simulations were performed by consecutive simulations with $2\times 10^8$ time steps.
At the start of a simulation the center of mass  and momentum components were adjusted to zero, and by the end of a
simulation the round-off errors had changed   the center of mass  components from $\approx 10^{-18}$ to $\approx 10^{-16}$.
The components of the momentum was changed
with the same factor. These tiny adjustments have, however, no effect on the stability, nor on the orbits of the planetary systems.

  Without any approximations the dynamics of a planetary systems is time demanding because
   one needs a rather small time increment $\delta t$ to obtain the orbit of a planet  at Perihelion accurately,
   and the stability of the planetary system is given by
    its long-time behaviour, which together implies that it is necessary to performs  billions of time steps in order to
     obtain the long-time behaviour of a planetary system. The present calculations are for $n= 1.8 \times 10^9$ time steps ($t=4.5\times 10^6$).
      Another complication is that the computational time without some approximations  varies with the number $N$ of objects as $\propto N^2$.
       It is, however, straight forward to implement different kind of time consuming approximations used
        in MD \cite{Hockney1974,Toxdyr} whereby the computational time
	 varies proportional to  $\approx N$.
\end{document}